\documentclass[sigconf]{acmart}
\usepackage{pgfplots}
\usepackage{subfigure}
\usetikzlibrary{external}

\AtBeginDocument{%
  \providecommand\BibTeX{{%
    \normalfont B\kern-0.5em{\scshape i\kern-0.25em b}\kern-0.8em\TeX}}}

\copyrightyear{2021}
\acmYear{2021}
\setcopyright{acmlicensed}
\acmConference[AIES '21] {Proceedings of the 2021 AAAI/ACM Conference on AI, Ethics, and Society}{May 19--21, 2021}{Virtual Event, USA.}
\acmBooktitle{Proceedings of the 2021 AAAI/ACM Conference on AI, Ethics, and Society (AIES '21), May 19--21, 2021, Virtual Event, USA}
\acmPrice{15.00}
\acmISBN{978-1-4503-8473-5/21/05}
\acmDOI{10.1145/3461702.3462563}

\begin{document}
\fancyhead{}

\title{The Grey Hoodie Project:\\ Big Tobacco, Big Tech, and the Threat on Academic Integrity}

\author{Mohamed Abdalla}
\email{msa@cs.toronto.edu}
\affiliation{%
  \institution{Centre for Ethics \& Department of Computer Science \\University of Toronto }
  \country{Canada}
}

\author{Moustafa Abdalla}
\email{moustafa\_abdalla@hms.harvard.edu}
\affiliation{%
  \institution{Harvard Medical School}
  \city{Cambridge}
  \country{The United States of America}}

\renewcommand{\shortauthors}{Abdalla and Abdalla}

\begin{abstract}
  As governmental bodies rely on academics' expert advice to shape policy regarding Artificial Intelligence, it is important that these academics not have conflicts of interests that may cloud or bias their judgement. Our work explores how Big Tech can actively distort the academic landscape to suit its needs. By comparing the well-studied actions of another industry (Big Tobacco) to the current actions of Big Tech we see similar strategies employed by both industries. These strategies enable either industry to sway and influence academic and public discourse. We examine the funding of academic research as a tool used by Big Tech to put forward a socially responsible public image, influence events hosted by and decisions made by funded universities, influence the research questions and plans of individual scientists, and discover receptive academics who can be leveraged. We demonstrate how Big Tech can affect academia from the institutional level down to individual researchers. 
Thus, we believe that it is vital, particularly for universities and other institutions of higher learning, to discuss the appropriateness and the tradeoffs of accepting funding from Big Tech, and what limitations or conditions should be put in place.
\end{abstract}


\keywords{research funding; conflict of interest}

\maketitle

\section{Introduction}
Imagine if, in mid-December of 2019, over 10,000 health policy researchers made the yearly pilgrimage to the largest international health policy conference in the world. Among the many topics discussed at this hypothetical conference was how to best deal with the negative effects of increased tobacco usage (e.g., tobacco related morbidity). Imagine if many of the speakers who graced the stage were funded by Big Tobacco. Imagine if the conference itself was largely funded by Big Tobacco.

Would academics in the field of public health accept this? Today, most would find such a situation inconceivable --- given the clear conflict of interest. In alignment with Article 5.3 of the WHO Framework Convention of Tobacco Control \cite{noauthor_who_2003}, policy makers would not look towards these speakers for advice regarding health policy. Anything said at this venue regarding the effects of smoking on public health would be met with skepticism and distrust. The negative effect of Big Tobacco's money on research quality has been widely reported, and it is commonly accepted that private-interest funding  biases research \cite{barnes_scientific_1997,brownell_perils_2009,cohen_institutional_1999,smith_public_2016}.

However, this is exactly what is happening in the field of machine learning. Replace ``health policy'' with ``machine learning'', ``effects of increased tobacco usage'' with ``ethical concerns of increased AI deployment'', and ``Big Tobacco'' with ``Big Tech'' and you get what is a significant, ongoing conflict of interest in academia. Yet this is largely regarded as a non-issue by many of those in the field.

In this work, we explore the extent to which large technology corporations (i.e., Big Tech) are involved in and leading the ongoing discussions regarding the ethics of AI in academic settings. By drawing upon historic examples of industry interference in academia and comparing these examples with the behavior of Big Tech, we demonstrate that there is cause for concern regarding the integrity of current research, and that academia must take steps to ensure the integrity and impartiality of future research.

\section{Defining Big Tech}

It is difficult to succinctly define which companies are or aren't ``Big Tech''. In this piece, as we discuss the ethics of AI, we focused on companies that are both large technology corporations and heavily involved in the ethics of AI/fairness literature or lobbying governments about such matters. There are some companies that are undoubtedly ``Big Tech'' (e.g., Google, Facebook) but the distinction is not clear for all companies (e.g., Yahoo (is it still big enough?), Disney (is it a technology company?)). The final list of companies was arrived at through multiple discussions and a final vote with various non-author parties. Our list is \textbf{purposefully more conservative} than most would be. By being more conservative, all of the conclusions arrived at in the paper would still hold (if not be strengthened) by selecting any additional companies.

In this paper, the following companies were considered ``Big Tech'': Google, Amazon, Facebook, Microsoft, Apple, Nvidia, Intel, IBM, Huawei, Samsung, Uber, Alibaba, Element AI, OpenAI.

The following companies were not considered ``Big Tech'': Disney, Autodesk, Pixar, Adobe, Polaroid, Pfizer, Sony, Oracle, Cisco, Netflix, Yahoo, VMWare, Activision, Pintrest, Yahoo.

\section{Motivating the analogy: Big Tobacco's and Big Tech's Playbook}

In this section, we will explore the histories of Big Tobacco and Big Tech. We see that both industries' increased funding of academia was as a reaction to increasingly unfavorable public opinion and an increased threat of legislation. The rest of the paper will explore the actions of Big Tech by drawing analogies to the actions of Big Tobacco.

We note that the analogy between Big Tobacco and Big Tech is not perfect. The analogy to Big Tobacco is intended to serve two purposes: 1) to provide a historical example with a rich literature to which we can compare current actions (i.e., help us know what to look for), and 2) to leverage the negative gut reaction to Big Tobacco's funding of academia to enable a more critical examination of Big Tech. For example, when positing that peer-review could help us address the issues raised by conflicts of interests (as has been done by some readers), we urge the reader to consider the same suggestion with respect to Big Tobacco; the fields of life science, biology, and public health have peer-review systems but did not view the existence of peer-review as a solution. The exact limitations of this analogy are covered in discussion.

\subsection{Big Tobacco}

In 1954, Big Tobacco was facing a decline in public opinion as demonstrated  and accompanied by the first ever decrease in demand of their product following the great depression \cite{brownell_perils_2009}. Just two years prior in Reader's Digest (which was a leading source of medical information for the general public), an article entitled ``Cancer by the carton'' was published discussing the link between smoking and lung cancer as presented by recent scientific studies \cite{norr_cancer_1952}.  While Big Tobacco internally acknowledged the conclusions drawn by these studies \cite{daube_towards_2017,glantz_truth_2000}, the threat to share value once the public was presented with this information was too large to leave unaddressed \cite{bates_tobacco_2004}. In response, Big Tobacco released a public letter titled ``A Frank Statement'' \cite{tobacco_industry_research_frank_1954}.

``A Frank Statement to Cigarette Smokers'' was a whole-page advertisement run by Big Tobacco in 1954 in over 400 newspapers reaching an estimated 43 million readers \cite{brownell_perils_2009,warner_tobacco_1991}. Signed by various presidents of Big Tobacco, the contents of the statement claimed  they ``accept an interest in people's health as a basic responsibility, paramount to every other consideration in our business'' and that Big Tobacco ``always have and always will cooperate closely with those whose task it is to safeguard the public's health'' \cite{tobacco_industry_research_frank_1954}. This public relations campaign (run by public relations firm  Hill \& Knowlton \cite{bero_lawyer_1995}) was part of a larger plan designed to both portray Big Tobacco as friendly corporations looking out for their consumers and purposefully sow doubt into the scientific research, which was showing conclusive links between smoking and lung cancer \cite{brandt_inventing_2012}.

An important part of Big Tobacco's plan was to ``cooperate closely with those whose task it is to safeguard the public's health'' \cite{tobacco_industry_research_frank_1954}. This included the creation of the Tobacco Industry Research Committee (TIRC), later renamed to Council for Tobacco Research (CTR) in 1964 \cite{warner_tobacco_1991}. The stated purpose of this Council was to ``to provide financial support for research by independent scientists into tobacco use and health'' \cite{warner_tobacco_1991}. A statement by the CTR published in 1986 would boast ``support of independent research is in excess of \$130 million and has resulted in publication of nearly 2,600 scientific papers, with eminent scientists thinking that questions relating to smoking and health were unresolved and the tobacco industry will make new commitments to help seek answers to those questions'' \cite{warner_tobacco_1991}. While the presented numbers are factually true, the underlying motivation behind such funding remained hidden until uncovered by litigation in 1998 \cite{bero_lawyer_1995,hurt_open_2009,malone_tobacco_2000}.

\subsection{Big Tech}

Just like Big Tobacco, Big Tech was starting to lose its luster (a trend which started in the second half of the 2010s \cite{doherty_americans_2019}). Public opinion of these large technology corporations was starting to sour, as their image shifted from savior-like figures to traditional self-seeking corporations \cite{doherty_americans_2019}. This decline in opinion was highlighted when it came to light that Facebook's platform was used by foreign agents to influence the 2016 US presidential election \cite{adams_facebooks_2018}. 

In Mark Zuckerberg's opening remarks to the US Congress, he stated ``it's clear now that we didn't do enough. We didn't focus enough on preventing abuse and thinking through how people could use these tools to do harm as well'' \cite{washington_post_transcript_2018}, and that Facebook was going to take their responsibility more seriously from now on. His opening statement is analogous to ``A Frank Statement'', failing to recall how leaked internal emails stated that they were aware of companies breaking Facebook's scraping policy, explicitly naming Cambridge Analytica \cite{mccarthy_facebooks_2019}. It also failed to mention how this was not the first, rather one of many, apologies made by the CEO to the public for negative (often purposeful) decisions that were later discovered by the public \cite{mccracken_brief_2018}.

Just like Big Tobacco, in response to a worsening public image, Big Tech had started to fund various institutions and causes to ``ensure the ethical development of AI'' \cite{european_commission_artificial_2019}, and to focus on ``responsible development'' \cite{walker_external_2019}. Facebook promised its ``commitment to the ethical development and deployment of AI'' \cite{facebook_engineering_ai_2018}. Google published its best practices for the ``ethical'' development of AI \cite{google_ai_artificial_2018}. Microsoft has claimed to be developing an ethical checklist \cite{boyle_microsoft_2019}, a claim that has recently been called into question \cite{todd_microsoft_2019}. Amazon co-sponsored, alongside the National Science Foundation, a \$20 million program on ``fairness in AI'' \cite{romano_amazons_2019}. In addition to these initiatives, Big Tech had been busy funding and initiating centers, which study the impact of their work on society. Big Tech's response to public criticism is similar to Big Tobacco's response: pump vast sums of money into these causes. As such, we must purposefully approach such contributions with caution, and make sure to study and understand the underlying motivations, interests (including financial interests), and conflicts of interest (perceived or actual).

\section{Methodology}

\begin{figure*}[ht]
  \centering
  \includegraphics[width=320pt]{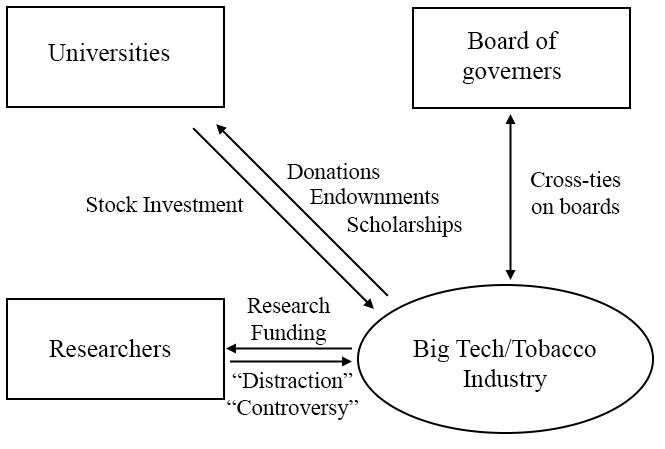}
    \caption{\label{fig:f1} Various ways tobacco industry money can find its way into academia, (recreated from \citet{cohen_institutional_1999}).}
\end{figure*}

Big Tobacco's investment in academic institutions, Figure \ref{fig:f1} (recreated from \cite{cohen_institutional_1999}), helped (and continues to help \cite{daube_towards_2017,waa_foundation_2020}) advance their business in numerous covert ways. 

From the literature on Big Tobacco, we observe four main goals driving investment into academia. For Big Tobacco, funding research in the academy serves to:
\begin{itemize}
    \item Reinvent itself in the public image as socially responsible;
    \item Influence the events and decisions made by funded universities;
    \item Influence the research questions and plans of individual scientists;
    \item Discover receptive academics who can be leveraged.
\end{itemize}

In this work, we explore whether Big Tech's funding of research in the academy can serve the same purpose (i.e., help them achieve these four goals). We do this by drawing 1:1 comparisons between Big Tobacco and Big Tech's actions for each of the four main goals. It is important to note that we can only see the intentions for Big Tobacco's actions because of the wealth of information revealed by litigation. Therefore this paper does not make the claim that Big Tech is intentionally attempting to influence academia as we cannot prove this claim. Rather we believe that industry funding warps academia regardless of intentionality due to perverse incentives.

\textbf{Ironically, Google adopts our view that industry funding warps research  \cite{miller_responding_2017}}. In their blog, Google questions the anti-Google work done by ``Campaign for Accountability'' \cite{noauthor_google_2017} solely because the non-for-profit is funded by Oracle which has incentive to hurt Google \cite{miller_responding_2017}. Abstracting the proper nouns: they are questioning work done by seemingly external researchers because the researchers are funded by a tech company that has external (profit) incentives.

\section{Reinvent itself in the public image as socially responsible}

\subsection{Big Tobacco}

Big Tobacco created its funding agencies in a seemingly impartial manner. The CTR was advised by various distinguished scientists who served on its scientific advisory board \cite{brandt_inventing_2012}. During its existence, hundreds of millions of dollars were provided to independent investigators at academic institutions across the US and abroad \cite{warner_tobacco_1991}. There is no doubt that a considerable amount of quality research was done as a result of this funding. However, the majority of the funding provided by the CTR went to research that was unrelated to the health effects of tobacco use \cite{bloch_tobacco_1994}. This contradicts the stated mission of ``tobacco industry will make new commitments to help seek answers to those questions [i.e., the effects of tobacco usage on health]''.

Why is this the case? Those responsible for funding, usually lawyers instead of scientists \cite{bero_lawyer_1995}, ``would simply refuse to fund any proposals that acknowledged that nicotine is addictive or that smoking is dangerous'' \cite{bero_lawyer_1995}. Furthermore, they sought out and funded projects that would shift the blame for lung cancer away from tobacco to other sources (e.g., birds as pets \cite{cunningham_smoke_1996}) \cite{bero_lawyer_1995}.

However, the purpose in funding so many projects was to use the act of funding as proof of social responsibility. This pretense of social responsibility was presented to juries in multiple cases, such as a cigarette product liability trial in 1990 Mississippi, during which a list of all the universities and medical schools supported by CTR grants was presented to jurors \cite{bloch_tobacco_1994}.

\subsection{Big Tech}

Just like Big Tobacco, lawyers and public relations are involved in plotting the research direction and the tone of research done at these companies (including those with external collaborators in academia) \cite{pos_tone_2020}. Leaked internal documents from Google, uncover PR directing ``its scientists to `strike a positive tone' in their research'' \cite{pos_tone_2020}.

Similar to how Big Tobacco created the TIRC to provide financial support to independent scientists to study tobacco and its influences on health, Big Tech has funded various similar institutions. Founded in 2016 by Google, Microsoft, Facebook, IBM, and Amazon among others, the ``Partnership on AI [henceforth: PAI] to Benefit People and Society'' was ``established to study and formulate best practices on AI technologies, [... and study]  AI and its influences on people and society'' \cite{partnership_on_ai_httpswwwpartnershiponaiorg_nodate}. Unfortunately, non-Big Tech members of this partnership realized that neither the ``ACLU nor MIT nor any other nonprofit has any power in this partnership'', leading members to conclude ``PAI's association with ACLU, MIT and other academic/non-profit institutions practically ends up serving a legitimating function'' \cite{ochigame_how_2019}. More recently, Access Now (a human rights organization), recently left PAI stating that they ``did not find that PAI influenced or changed the attitude of member companies or encouraged them to respond to or consult with civil society on a systematic basis'' \cite{access_now_2020}.

In addition to this joint partnership where industry interests prevail over public interest, each company has also been working on its own individual PR campaign. Google, in response to growing employee and public concern regarding its collaboration with the US military and Pentagon, created ``an external advisory council'' \cite{google_ai_artificial_2018}, which was later disbanded after a scandal surrounding one of its proposed members. Microsoft has created an internal committee to ``ensure our AI platform and experience efforts are deeply grounded within Microsoft's core values'' \cite{nadella_satya_2018}. In response to the Cambridge Analytica scandal, Zuckerberg promised that Facebook will create ``an independent group to study the abuse of social media in elections'', and that members will ``be independent academics, and Facebook [will have] no prior publishing control'' \cite{washington_post_transcript_2018}. More recently, in light of growing concerns regarding ``fake news'' and disinformation campaigns, Facebook announced it was giving away \$300 million in grants to support journalism, to the ire of some academics \cite{ingram_facebook_2018,ingram_facebook_2019}.

While these efforts are likely to result in some tangible good (as did research funded by Big Tobacco), these efforts will be limited by the profit motives of corporations \cite{pos_tone_2020}. Just as Big Tobacco leveraged research funding to avoid legal responsibility, Big Tech has used a similar line of argument to avoid scrutiny, demonstrated by Zuckerberg's usage of ``independent academics'' to congress, Google's boasting of ``releasing more than [75 or 200] research papers on topics in responsible AI [2019 or 2020]'' \cite{dean_responsible_2019,walker_responsible_2020}.

\section{Influence the events and decisions made by funded universities}

\subsection{Big Tobacco}

Positive PR was not the only motivating factor behind providing money to institutions. Evidence has shown that Big Tobacco gains undeserved influence in the decision making process of universities that are dependent on them for money \cite{landman_tobacco_2009,noauthor_preventing_2002}.

 Looking at a single public university, the University of Toronto (UofT), Imperial Tobacco withheld its (previously consistent) funding from the annual conference at the University of Toronto's Faculty of Law as retribution for the fact that UofT law students were influential in having criminal charges be laid against Shoppers Drug Mart for selling tobacco to a minor \cite{cunningham_smoke_1996}.  

Other, more subtle, effects of Big Tobacco's influence on academia's decision making is that of delayed decisions or institutional inaction on tobacco control issues \cite{cohen_institutional_1999}. While this can be achieved through funding threats, it is also possible through the planting or recruitment of friendly actors in academia. Examples of this include how the former President and Dean of law at the UofT, Robert Prichard, was a director of Imasco (a large tobacco company) \cite{cohen_institutional_1999,cunningham_smoke_1996}. Additionally, Robert Parker who was the president and chief spokesperson for the Canadian Tobacco Manufacturers' Council, was also on the Board of the Foundation of Women's College Hospital, a teaching hospital also affiliated with UofT \cite{cohen_institutional_1999}. The network of such placements, which have been documented in universities across many countries \cite{cunningham_smoke_1996,gruning_tobacco_2006}, demonstrates how a university's decisions could be affected by conflicts of interests.

Additionally, events sponsored by Big Tobacco (e.g., symposiums held about second hand smoking) have been shown to be skewed and of poorer quality compared to events not sponsored by Big Tobacco \cite{barnes_scientific_1997,bero_sponsored_1994} but still are cited by Big Tobacco when supporting its interests \cite{barnes_scientific_1997,bero_sponsored_1994}.

\subsection{Big Tech}

Similar to Big Tobacco, positive PR is not the only motivating factor for Big Tech when providing funding for institutions. In addition to academic innovation and research helping advance industrial products, this funding also gives Big Tech a strong voice in what happens in conferences and in academia.

The top machine learning conference NeurIPS has had at least two Big Tech sponsors at the highest tier of funding since 2015. In recent years, the number of Big Tech companies at the highest tier of funding has exceeded five\footnote{The sponsorship info for NeurIPS was obtained from the ``Sponsor Information for NeurIPS 20XX'' page for each conference (e.g., \url{https://nips.cc/Sponsors/sponsorinfo}). We only counted Big Tech companies as sponsors if they were sponsoring at the highest possible level for the respective year.}. When considering workshops relating to ethics or fairness\footnote{For each workshop, we gathered the list of organizers on the workshop's website and searched the web for author affiliations (both past and present).}, all but one have at least one organizer who is affiliated or was recently affiliated with Big Tech. For example, there was a workshop about ``Responsible and Reproducible AI'' sponsored solely by Facebook. From 2015 to 2019, the only ethics-related workshop at NeurIPS that did not have at least one organizer belonging to Big Tech, was 2018's ``Robust AI in Financial Services'' workshop (which instead featured 4 heads of AI branches at big banks). 

Such a conflict of interest persists even when considering conferences dedicated to examining the societal effects of technology. For example, FAccT (previously known as FAT*) has never had a year without Big Tech funding: Google (3/3 years), Microsoft (2/3 years), Facebook (2/3 years)\footnote{The sponsorship info for FAccT was obtained from the ``Sponsors and Supporters'' page for each conference (e.g., \url{https://facctconference.org/2020/sponsorship.html})}. While the conference organizers provide a ``Statement Regarding Sponsorship and Financial Support'', it's not clear how effective such a policy is at preventing the unconscious biasing of attendees and researchers. That is, despite research in other fields clearly demonstrating that industry funding negatively impacts work and subconsciously biases researchers  \cite{goldberg2019shadows, marks2020lessons, new_sci_2015,owram2004managing}, many organizers and academics in computer science believe that this is not cause for concern without offering evidence to the contrary.

In public health policy, disclosure of conflicts of interests is simply seen as a mechanism to indicate the existence of a problem. We believe, as argued by \citet{goldberg2019shadows} ``rather than disclosure and management, the ethically paramount intervention targeted  against  behavior  of  partiality  flowing  from [conflict of interests]  is  the  idea  of  sequestration.  Sequestration  refers to the idea of eliminating or at least severely curtailing relationships between commercial industries and [...] professionals''.

Furthermore, there is work \cite{foroohar_dont_2019} which demonstrates how this funding further purports Big Tech's views and what solutions are and are not acceptable \cite{metcalf_owning_2019}. By controlling the agenda of such workshops, Big Tech controls the discussions, and can shift the types of questions being asked and the direction of the discussion. A clear example of this was when, ``[as] part of a campaign by Google executives to shift the antitrust conversation'', Google sponsored and planned a conference to influence policy makers going so far as to invite a ``token Google critic, capable of giving some semblance of balance'' \cite{foroohar_dont_2019}. Another example was the ``Workshop on Federated Learning and Analytics'' organized by Google researchers. Distinguishing between little ``p'' privacy from big ``P'' Privacy (the former being protecting data from adversaries (i.e., confidentiality) and the latter focusing on privacy as a human right with societal value), the workshop focused more on privacy over Privacy \cite{fed_priv_2019}. 

Just like Big Tobacco, Big Tech has been busy building relationships with leaders in academia. Focusing again on the University of Toronto, the Vector Institute, has/had as faculty members a Vice President of Google and the heads of various companies' AI branches such as Uber, and Nvidia. And although the institute is mainly funded by governments (about one third comes from industry \cite{the_canadian_press_federal_2017}), they have largely remained quiet regarding any of the ethical issues caused by those providing them funding. This is not necessarily because of their funders, but it would be unreasonable to assume that the risk of losing a third of one's funding wouldn't impact what the institute does in the public sphere. This is despite having some of the most renowned researchers on ``fairness'' in the field. In reality, fairness is relegated simply to a mathematical problem. This formulation of existing issues is in-line with the dominant ``logics of Big Tech'' \cite{metcalf_owning_2019, fazelpour2020algorithmic} which fails to consider the many questions and concerns raised by those outside of Big Tech. 

\section{Influence the research questions and plans of individual scientists}

\subsection{Big Tobacco}

CTR purposefully funded many projects not related to studying the direct effects of increased tobacco usage on health outcomes. Through the allocation of its funds, it directly and indirectly impacted research questions  and the direction of research when it came to the health effects of smoking \cite{bero_lawyer_1995}. First and foremost, Big Tobacco actively sought out to fund any research that attempted to shift the blame from tobacco to other sources \cite{bero_lawyer_1995,bloch_tobacco_1994}. When this was not possible, Big Tobacco opted to steer funds from exploring the health effects of tobacco to studying the basic science of cancer \cite{bero_lawyer_1995}. By dropping the tobacco link, the research was viewed as less threatening and therefore ``fundable'', and, in other words, distracting scientists and the public by sowing seeds of confusion and discord in the public and scientific community.

Other actions included threatening to ``take out ads [...] that point out the flaws of studies'' in an attempt to shame scientists and make working in the area against Big Tobacco a more difficult endeavor with no room for mistakes \cite{landman_tobacco_2009, philip_morris_draft_1995}. Phillip Morris and RJ Reynolds (large tobacco companies) also worked with elected politicians to block the funding of scientists with opposing viewpoints: ensuring ``that the labor-HHS (US Department of Health and Human Services) Appropriations continuing resolution will include language to prohibit funding for Glantz [a scientist opposing Big Tobacco]'' \cite{landman_tobacco_2009,philip_morris_national_1995}. 

As most researchers have to seek out grants from funding bodies to perform their research, it is quite likely that they would seek funding from Big Tobacco or institutions under the sway of Big Tobacco (such as NCI and HHS). To increase the chances of approval, it would make sense that researchers would make changes to the types of questions they would explore, as it has been made clear that Big Tobacco would not be funding certain questions.  

\subsection{Big Tech}
\label{sec:incentives}

Just as many Big Tobacco-funded projects lead to tangible advancements in science and improved the lives of people, the same can be said for the majority of the work funded by Big Tech. As is the case with Big Tobacco, there is evidence that the types of questions being asked, the types of projects being funded, and the types of answers being provided are influenced by Big Tech.

An especially egregious example demonstrating how Big Tech attempts to influence the research questions and plans of scientists comes from leaked internal documents from Google, which tells ``its scientists to `strike a positive tone' in their research'' \cite{pos_tone_2020}. Just like tobacco companies, for research surrounding certain topics (e.g., effects of AI), researchers must consult with ``policy and public relations teams'' \cite{pos_tone_2020}.  At the very least, the public relation goals of the company influences academics who collaborate directly with industry researchers. This should be cause for concern, as the processes for approving industry grants to academics in universities may undergo similar public relations clearances thus suppressing work in the academy. 

\subsection*{Faculty Funding}
A critical way Big Tech gains influence over AI ethicists, is by acting as a pseudo-granting body. That is, by providing a large amount of money to researchers, Big Tech is able to decide what will and won't be researched. We show that a majority (58\%) of AI ethics faculty are looking to Big Tech for money meaning that Big Tech is able to influence what they work on. This is because, to bring in research funding, faculty will be pressured to modify their work to be more amenable to the views of Big Tech. This influence can occur even without the explicit intention of manipulation, if those applying for awards and those deciding who deserve funding do not share the same underlying views of what ethics is or how it ``should be solved''.

To demonstrate the scope of Big Tech funding in academia, we explored the funding of tenure-track research faculty in the computer science department at 4 R1  (top PhD granting) universities: Massachusetts Institute of Technology (MIT), UofT, Stanford, and Berkeley. We show that 52\% (77/149) of faculty with known funding sources (29\% of total) have been directly funded by Big Tech, Table \ref{tab1}. Expanding the criteria to include funding at any stage of career (i.e., PhD funding) as well as previous work experience, we find 84\% (125/148) of faculty with known funding sources (47\% of total) have received financial compensation by Big Tech, Table \ref{tab4}. 

Both these percentages rise when we limit our analysis to faculty who have published at least one ethics or ``fairness'' paper between January 2015 and April 2020. With this criteria, we find that 58\% (19/33) of faculty with known funding sources (39\% total) have, at one point, been directly funded by Big Tech, Table \ref{tab3}. Expanding the funding criteria to include graduate funding as well as previous work experience, we note that 97\% (32/33) of faculty with known funding sources (65\% total) have received financial compensation by Big Tech, Table \ref{tab6}.

\subsubsection{Methodology}
Of the 4 R1 institutions (UofT, MIT, Stanford and Berkeley) we chose to study, two are private institutions (MIT and Stanford), while two are public institutions (UofT and Berkeley). Two of these institutions are on the eastern seaboard (UofT and MIT) and two are on the western seaboard (Stanford and Berkeley). 
For each of these universities, we gathered a list of professors from their universities' faculty listing for the computer science department:
\begin{itemize}
    \item UofT: \url{https://web.cs.toronto.edu/contact-us/faculty-directory}
    \item MIT: \url{https://www.eecs.mit.edu/people/faculty-advisors/35}
    \item Stanford: \url{https://cs.stanford.edu/directory/faculty}
    \item Berkeley: \url{https://www2.eecs.berkeley.edu/Faculty/Lists/CS/faculty.html}
\end{itemize}

We removed all professors who were not both research stream and tenure track or those who where emeritus.

For each professor, we assessed them according to the following categories:
\begin{itemize}
    \item ``Works on AI?'': This was scraped from the department's page where each faculties' interests were listed.
    \item ``Works on Ethics of AI?'': This was defined as having at least 1 ethics of AI/societal impacts of AI paper published from January 2015 to April 2020.
    \item ``Faculty funding from Big Tech'': Has this faculty won any awards or grants from any of the companies classified as Big Tech? This field could be responded to with one of Yes, No or Unknown. Unknown was used to represent faculty who did not have enough information published on their website to allow us to make a conclusion. Of course, it may be possible that those classified as ``No'' simply chose not to list such awards on their personal websites, but we chose to treat published CVs as fully comprehensive.
    \item ``Graduate funding from Big Tech'': Was any portion of this faculty's graduate education funded by Big Tech? This includes PhD fellowships and post-docs. Like before, this field could be responded to with one of Yes, No or Unknown.
    \item ``Employed by Big Tech'': Did this faculty at any time work for any Big Tech company? This includes roles as visiting researcher, consultant, and internships. Like before, this field could be responded to with one of Yes, No or Unknown.
\end{itemize}

\subsubsection*{Faculty Funding -- Analysis}
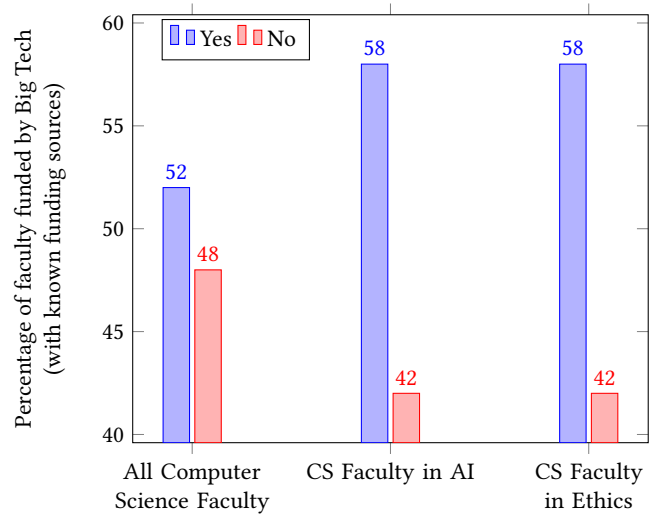
\begin{figure}[ht]
    \centering
\begin{tikzpicture}  
\begin{axis}  
[  
    ybar, 
    enlargelimits=0.15,
    legend style={at={(0.2,0.99)}, 
      anchor=north,legend columns=-1},     
    ylabel style = {align=center},
    ylabel={Percentage of faculty funded by Big Tech\\ (with known funding sources)}, 
    xticklabel style={text width=70, align=center},
    symbolic x coords={All Computer Science Faculty, CS Faculty in AI, CS Faculty in Ethics},  
    xtick=data,  
    nodes near coords,  
    nodes near coords align={vertical},  
    ]  
\addplot coordinates {(All Computer Science Faculty, 52) (CS Faculty in AI, 58) (CS Faculty in Ethics, 58)}; 
\addplot coordinates {(All Computer Science Faculty, 48) (CS Faculty in AI, 42) (CS Faculty in Ethics, 42)};  
\legend{Yes, No}  
  
\end{axis}  
\end{tikzpicture}
    \caption{Bar chart presenting the percentage of computer science faculty members who received (at any point in their career) direct funding from Big Tech, stratified by different areas of specialization. NB: The axis intentionally does not start at zero.}
    \label{fig:percent_funded}
\end{figure}

Our initial analysis explores direct funding of research by Big Tech. More specifically, we use our collected data to answer the question: ``Has this faculty won any awards, grants, or similar awards from any of the companies classified as Big Tech?'' Results for faculty with known funding sources is plotted in Figure \ref{fig:percent_funded}, with detailed results in Tables \ref{tab1},\ref{tab2},\ref{tab3}. Although we compare all computer science professors against those in ``Ethics of AI'', we present faculty working in the area of AI to present what may be a large confounding factor (i.e., Big Tech's growing interest in AI regardless of its ethics).

\begin{table}[ht]
\caption{The number of computer science faculty who have won funding, grants, or similar awards from any of the companies classified as Big Tech. The results are also stratified by school.}
\label{tab1}
\begin{tabular}{l|c|c|c}
\textbf{}               & \textbf{Yes} & \textbf{No} & \textbf{Unknown} \\ \hline
\textbf{All Professors} & 77           & 72          & 118              \\
\textbf{UofT}           & 17           & 15          & 30               \\
\textbf{MIT}            & 22           & 14          & 30               \\
\textbf{Stanford}       & 20           & 16          & 25               \\
\textbf{Berkeley}       & 18           & 27          & 33              
\end{tabular}
\end{table}

\begin{table}[ht]
\caption{The number of computer science faculty working on AI who have won funding, grants, or similar awards from any of the companies classified as Big Tech. The results are also stratified by school.}
\label{tab2}
\begin{tabular}{l|c|c|c}
\textbf{}               & \textbf{Yes} & \textbf{No} & \textbf{Unknown} \\ \hline
\textbf{All Professors} & 48           & 35          & 52              \\
\textbf{UofT}           & 12           & 10          & 11               \\
\textbf{MIT}            & 14           & 7          & 19               \\
\textbf{Stanford}       & 11           & 8          & 10               \\
\textbf{Berkeley}       & 11           & 10          & 12              
\end{tabular}
\end{table}

\begin{table}[ht]
\caption{The number of computer science faculty who have at least 1 ``Ethics of AI'' publication who have won funding, grants, or similar awards from any of the companies classified as Big Tech. The results are also stratified by school.}
\label{tab3}
\begin{tabular}{l|c|c|c}
\textbf{}               & \textbf{Yes} & \textbf{No} & \textbf{Unknown} \\ \hline
\textbf{All Professors} & 19           & 14          & 16              \\
\textbf{UofT}           & 4           & 3          & 3               \\
\textbf{MIT}            & 2           & 2          & 5               \\
\textbf{Stanford}       & 6           & 2          & 3               \\
\textbf{Berkeley}       & 7           & 7          & 5              
\end{tabular}
\end{table}

\subsubsection*{Faculty Association -- Analysis}

This analysis explores any direct financial relationship between faculty members and Big Tech (past or present). More specifically we use our collected data to answer the question: ``Has this faculty won any awards, grants, or similar awards from any of the companies classified as Big Tech?'' \textit{OR} ``Was any portion of this faculty's graduate education funded by Big Tech?'' \textit{OR} ``Did this faculty at any time work for any Big Tech company?'' Although we compare all computer science professors against those in ``Ethics of AI'', we also present faculty working on AI to present what may be a large confounding factor (i.e., Big Tech's growing interest in AI regardless of its ethics).

The purpose of such collection is not to imply that any past financial relationship would necessarily have a detrimental and conscious impact on the research of the scholar (i.e., we are not implying that graduate funding sources will consciously bias a professor's views 10 years later). Rather, we believe that given the dominant views of ethics by Big Tech, repeated exposure to such views (in a positive setting) is likely to result in increased adoption and be a means of subconscious influence.

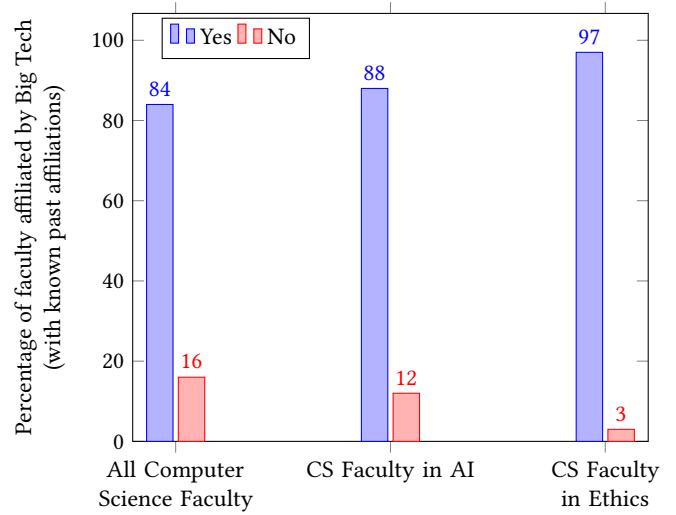
\begin{figure}[ht]
    \centering
\begin{tikzpicture}  
\begin{axis}  
[  
    ybar, 
    ymin=0,
    legend style={at={(0.2,0.99)}, 
      anchor=north,legend columns=-1},     
    ylabel style={align=center},
    ylabel={Percentage of faculty affiliated by Big Tech\\ (with known past affiliations)}, 
    xticklabel style={text width=70, align=center},
    symbolic x coords={All Computer Science Faculty, CS Faculty in AI, CS Faculty in Ethics},  
    xtick=data,  
    nodes near coords,  
    nodes near coords align={vertical},  
    ]  
\addplot coordinates {(All Computer Science Faculty, 84) (CS Faculty in AI, 88) (CS Faculty in Ethics, 97)}; 
\addplot coordinates {(All Computer Science Faculty, 16) (CS Faculty in AI, 12) (CS Faculty in Ethics, 3)};  
\legend{Yes, No}  
  
\end{axis}  
\end{tikzpicture}
    \caption{Bar chart presenting the percentage of computer science faculty members who have at any point in their career received direct funding/awards from Big Tech or have been employed by Big Tech stratified by different areas of specialization.}
    \label{fig:percent_anything}
\end{figure}

\begin{table}[ht]
\caption{The number of computer science faculty who have had any financial association with Big Tech (e.g., won funding, were employees or contractors, etc.,). The results are also stratified by school.}
\label{tab4}
\begin{tabular}{l|c|c|c}
\textbf{}               & \textbf{Yes} & \textbf{No} & \textbf{Unknown} \\ \hline
\textbf{All Professors} & 125           & 23          & 119              \\
\textbf{UofT}           & 29           & 5          & 28               \\
\textbf{MIT}            & 31           & 4          & 31               \\
\textbf{Stanford}       & 32           & 5          & 24               \\
\textbf{Berkeley}       & 33           & 9          & 36              
\end{tabular}
\end{table}

\begin{table}[ht]
\caption{The number of computer science faculty working on AI who have had any financial association with Big Tech (e.g., won funding, were employees or contractors, etc.,). The results are also stratified by school.}
\label{tab5}
\begin{tabular}{l|c|c|c}
\textbf{}               & \textbf{Yes} & \textbf{No} & \textbf{Unknown} \\ \hline
\textbf{All Professors} & 75           & 10          & 50              \\
\textbf{UofT}           & 19           & 4          & 10               \\
\textbf{MIT}            & 19           & 1          & 20               \\
\textbf{Stanford}       & 9           & 0          & 2               \\
\textbf{Berkeley}       & 14           & 1          & 4              
\end{tabular}
\end{table}

\begin{table}[ht]
\caption{The number of computer science faculty with at least 1 ``Ethics of AI'' publication who have had any financial association with Big Tech (e.g., won funding, were employees or contractors, etc.,). The results are also stratified by school.}
\label{tab6}
\begin{tabular}{l|c|c|c}
\textbf{}               & \textbf{Yes} & \textbf{No} & \textbf{Unknown} \\ \hline
\textbf{All Professors} & 32           & 1          & 16              \\
\textbf{UofT}           & 7           & 0          & 3               \\
\textbf{MIT}            & 2           & 0          & 7               \\
\textbf{Stanford}       & 9           & 0          & 2               \\
\textbf{Berkeley}       & 14           & 1          & 4              
\end{tabular}
\end{table}

\subsubsection*{Funder -- Analysis}
Having shown the majority of faculty at these four school have at one point been directly funded by Big Tech, in this analysis we explore which corporations are directly responsible for the funding, Table \ref{tab7}. 

\begin{table}[ht]
\caption{Top 6 Big Tech companies which contribute directly to faculty research through grants, research awards, or similar means.}
\label{tab7}
\begin{tabular}{l|c||c|c|c|c}
\textbf{}          & \textbf{All} & \textbf{UofT} & \textbf{MIT} & \textbf{Stanford} & \textbf{Berkeley} \\ \hline
\textbf{Google}    & 44           & 8            & 13          &    10               &  13                 \\
\textbf{Microsoft} & 25           & 3            & 5           &      9             &     7              \\
\textbf{Amazon}    & 14           & 3            & 3           &       3            &     5              \\
\textbf{IBM}       & 13           & 3            & 3           &        3           &     5              \\
\textbf{Facebook}  & 13           & 3            & 3           &         4          &     3    \\
\textbf{Nvidia}  & 9           & 5            & 0           &             1      &       3  
\end{tabular}
\end{table}

\subsection*{Journal Author Funding and Association}
To demonstrate the influence of Big Tech funding in driving discussion regarding AI ethics and fairness in non-technical academia, we performed a systematic review of all articles published in the two leading non-technical journals: Nature and Science. 

We find that of the 59\% (10/17) of position papers ever published regarding the ethical/societal impact of AI has at least one author who was financially involved with one of these companies at one point in time (including faculty awards and former consulting gigs). 

\subsubsection{Methodology}
In both Nature and Science, we obtained all commentary or prospective pieces published after 2015 that returned as a result of the following search terms ($n=68$):
\begin{itemize}
    \item machine learning AND fairness,
    \item machine learning AND bias,
    \item machine learning AND ethics,
    \item machine learning AND racist,
    \item machine learning AND disparity,
\end{itemize}

We removed any articles that were not published by members of the academy or industry (i.e., journalists for the respective journals). We subsequently identified 17 papers of the remaining 51 articles as focusing on both AI and societal impacts (i.e. ``ethics'' or ``implications'' of AI and AI-related research). 

At a minimum, these statistics\footnote{The data for all analyses discussed in this paper can be requested from the authors.} demonstrate a clear conflict of interest between Big Tech and the research agendas of academics, regardless of intentions. As a result, it makes sense that much of the fairness work that exists holds the entrenched Big Tech view that ``social  problems  can  be addressed  through innovative technical solutions'' \cite{metcalf_owning_2019}. We do not claim that opposing viewpoints are not present in academia. However, opposing viewpoints are likely to comprise a minority proportion of the work and discussion presented at such workshops, conferences, and symposiums. 

\section{Discover receptive academics who can be leveraged}

\subsection{Big Tobacco}

Part of the strategy devised by Hill (of Hill \& Knowlton) leveraged skeptics within academia to sow doubt and foster distrust in the scientific findings \cite{brandt_inventing_2012}. These skeptics were solicited, given funding, and had their message amplified into the public discourse \cite{boyse_note_1988,brandt_inventing_2012,philip_morris_europe_organization_1987}. The result of such amplification resulted in new skeptics and the emboldening of existing ones -- something in line with Big Tobacco's goals. Based on memos released during litigation \cite{bero_lawyer_1995}, Big Tobacco's lawyers actively sought to discover academics whose research was sympathetic to their cause with the aim of funding any research that would allow them to claim that evidence regarding tobacco and lung cancer was inconclusive, such as research exploring if it was the keeping of birds as pets, as opposed smoking that increased the risk of lung disease \cite{bero_lawyer_1995,brandt_inventing_2012}.

In addition to these activities, funding was reserved for researchers who would be used to testify at legislative hearings in favor of Big Tobacco. In fact, there was a concentrated covert effort on behalf of Philip Morris International to identify European scientists with no previous connections to tobacco companies who could be potentially persuaded to testify on behalf of Big Tobacco against proposed regulation on second hand smoking \cite{boyse_note_1988}. This was part of the larger Whitecoat Project which resulted in infiltrations in governing officials, heads of academia, and editorial boards \cite{brandt_inventing_2012,landman_tobacco_2009,philip_morris_europe_organization_1987}.

\subsection{Big Tech}

Just as Big Tobacco leveraged its funding and initiatives to identify academics who would be receptive to industry positions and who, in turn, could be used to combat legislation and fight litigation, Big Tech leverages its power and structure in the same way. 

Google was noted to ``[groom] academic standard-bearers, prominent academics who will drive younger peers in a direction that is more favorable to the company'' \cite{foroohar_dont_2019}. In an article published by The Intercept, we discover that Eric Schmidt, previously of Google, was advised on which ``academic AI ethicists Schmidt's private foundation should fund'' \cite{ochigame_how_2019}. This is not a one-time occurrence either. Schmidt also inquired to Joichi Ito (formerly of MIT's Media Lab) if he ``should fund a certain professor who, like Ito, later served as an ``expert consultant'' to the Pentagon's innovation board'' \cite{ochigame_how_2019}. Another example of this recruitment is a professor at George Mason University, who had ``written academic research funded indirectly by Google, and criticized antitrust scrutiny of Google shortly before joining the Federal Trade Commission, after which the FTC dropped their antitrust suit'' \cite{foroohar_dont_2019,noauthor_google_2017}. Or consider the case where Schmidt cited a Google-funded paper when writing to congress without mentioning that the paper had been funded by Google \cite{noauthor_google_2017}. 

To demonstrate that this is not just the actions of high-level executives, consider the example of a former employee in the policy and public relations department at Google. The employee refutes Google's claim that ``Google’s collaborations with academic and research institutions are not driven by policy influence in any way'' \cite{will_knight_2020}, and presents an example where they, personally, intentionally leveraged a professor to influence policy \cite{will_fitz_2020,bruno_2015}. 

There is also investigative journalism that has uncovered examples of academics being funded by industry giants after presenting policy views \cite{wachter2017right} positive of industry positions despite numerous criticisms from other experts in the field \cite{williams_2019}. Following this funding, the academics released more policy position papers in line with industry positions without disclosing the financial connection \cite{williams_2019}. 

Such blatant and egregious interaction with academia harkens to Big Tobacco's Project Whitecoat. The name of our paper is an homage to Project Whitecoat: Project Grey Hoodie is referencing the buying out of technical academics. These connections are not fully exposed or available to the general public or the majority of academics and thus quite difficult to analyze because unlike Big Tobacco there has been no litigation to uncover the needed documents for analysis \cite{noauthor_preventing_2002}.

\section{Discussion and Conclusion}

The conflict of interest in academia caused by industry funding is a systemic issue which exists at the societal level. Therefore, we believe that effective solutions will have to come from policy (either governmental or institutional). It is important to stress that our examination of individual researchers is not meant to call their integrity into question. We think that the vast majority of work in academia is done by well-intentioned and skilled researchers. We restate, that this influence can occur without an explicit intent to manipulate, but simply through repeated interactions where one party takes substantial sums of money or spends a large amount of time in an environment with different goals/views.

There are various limitations to our work. First, the funding patterns we observed cannot be generalized to all academics. There is work that shows that there is a concentration of computing power and funding at ``elite'' institutions (the four schools we considered would be classified as such) \cite{ahmed2020democratization}.

In addition to this, the analogy between Big Tobacco and Big Tech is both polarizing and imperfect. First, unlike Big Tobacco, Big Tech is largely considered to have had a net positive impact on society. Discussion of the effects of increased AI deployment is more difficult to conduct as the outcomes of these algorithms are not so obvious when compared to the effects of increased tobacco usage (i.e., it's much easier to point to the death of an individual, than it is to demonstrate that Facebook's recommendation algorithms are hurting society/democracy \cite{statt_nick_2019}). This is especially the case when Big Tech is simply acting as an intermediary to other actions (e.g., providing facial recognition as a service (to the police) or computer vision algorithms as a service (to the military) versus actually performing policing using facial recognition). Further compounding this issue, is the near impossibility of doing critical work without Big Tech granting access to systems that they protect as trade secrets. This means that academics wishing to do good must expose themselves to conflicts of interests, though we believe with smart legislation these companies can be forced to allow unaffiliated academics to access and analyze their systems. While we acknowledge the limitations of the analogy, our usage of it is purposeful. Initial feedback to our paper (from those not familiar with how conflicts of interests work) was very defensive. While workshopping the paper, well-meaning academics could not understand how this funding could sway research direction (and believed that ``peer review'' was enough to protect us against influence, though it is unclear why as life-science/health policy/public health also have peer-review). We found that by priming an intuitive negative gut reaction to conflicts of interest by bringing up Big Tobacco the reader was more likely to critically engage with the ideas raised in the paper.

In this work, we believe we have shown that the interactions between academia and Big Tech especially regarding studying the impact of technology on society are eerily similar to those between academia and Big Tobacco in the late 20th century. The truly damning evidence of Big Tobacco's behavior only came to light after years of litigation \cite{hurt_open_2009}. However, the parallels between the public facing history of Big Tobacco's behavior and the current behavior of Big Tech should be a cause for concern\footnote{This parallel itself was made by a former executive of Facebook in reference to the design of their product. Former Facebook executive Tim Kendall's congressional testimony states that in the pursuit of profit, Facebook ``took a page from Big Tobacco's playbook'' \cite{kendall_congress_2018}.}. Having been a part of this narrative before, the academy is responsible for not allowing history to repeat itself once again. Rephrasing calls to action from the fight against Big Tobacco ``academic naiveté about [technology] companies' intentions is no longer excusable.  The extent of the [technology] companies'  manipulation [whether intentional or otherwise] needs  to  be  thoroughly [researched and] exposed'' \cite{yach_junking_2001}. 

There are many publicly proposed solutions to the societal problems caused by Big Tech from breaking up the companies to (and hopefully including) fixing the tax codes such that public institutions no longer need to rely on external funding for wages or to do their work. While we leave discussion regarding such in-depth solutions to a later work, we encourage the readers in academia to consider a stricter code of ethics and operation for AI-ethics research, separate from the traditional computer science department. Such a separation would permit academia-industry relationships for technical problems where such funding is likely more acceptable, while ensuring that our development of ethics remains free of influence from Big Tech money. 

We understand that it might not be possible (and some would argue undesirable) to completely divorce academia from Big Tech. However, financial independence should be a requirement for those claiming to study the effect of their technologies on our society.  Any change that is undertaken must be deliberate and structural in nature. However, in the meanwhile, here are a few steps that can be done right now to help decide future steps and questions:
\begin{itemize}
    \item Every researcher should be required to post their complete funding information online. Lack of information was the biggest stumbling block to analyzing funding sources of current academics and in turn the possible effect industry has on academia. Any and all historic affiliations should also be listed as to enable studying of research networks.
    \item Universities need to publish documents highlighting their position regarding the appropriateness of direct researcher funding from Big Tech. These documents should be created to answer questions such as: Should Big Tech be able to directly fund a researcher's work? Answering yes implies that Big Tech should have the ability to dictate what sort of questions public institutions should be looking at. While in certain technical problems this is appropriate, do the benefits outweigh the risks? Should the decision be made at the institution level or department by department? Maybe the answers depend on scientists' research area? What are some possible alternatives? Maybe industry funding can only be directed at national funding bodies or departments but not individual researchers?
    \item There needs to be discussion regarding the future of the ethics and fairness of the AI field and their dealings with industry. Is it permissible to seek external funding sources given the historical effects such funding has had on critical work? How do we ensure that their work is not co-opted by the industry to push agendas that are not agreeable with societal goals?
    \item Computer science as a field should explore how to actively court antagonistic thinkers. To undo the current concentration of influence in the ethics of AI, computer scientists will have to forcefully expose themselves to unpleasant opposition to their ideas and meet this opposition with an open and receptive mind.
\end{itemize}

None of these points can be effectively done in a vacuum. We believe that academics should learn from the experiences of other fields when dealing with corporate initiatives. 

\section{Acknowledgments}
The majority of Mohamed's funding comes from the Government of Canada (a Vanier scholarship), the Vector Institute, the University of Toronto, and the Centre for Ethics at the University of Toronto. The Vector institute is largely funded by public money but about one third comes from industry. Mohamed has previously interned at Google. Moustafa has no funding conflicts to report.

Special thanks is owed to LLana James and Raglan Maddox for their insight and feedback. We'd like to thank Saif Mohammad for his helpful discussions and pointing us to relevant news stories and Twitter threads with constructive criticism (and those providing constructive criticism in those threads). We are grateful for the reviewer' feedback from AIES and the Resistance Workshop at NeurIPS 2020.  We'd also like to thank Amelia Eaton for her copy-editing and feedback.
\bibliographystyle{ACM-Reference-Format}
\bibliography{paper}

\end{document}